\documentclass[prb, twocolumn, showpacs, amsmath, amssymb, amsfonts
]{revtex4}
\usepackage{graphicx}
\usepackage{graphics}
\usepackage{dcolumn}
\usepackage{bm}
\usepackage{amssymb}
\usepackage{amsmath}
\usepackage{amsfonts}
\usepackage{epsfig}

\newcommand {\bra} [1] {\langle #1 |}
\newcommand {\ket} [1] {| #1 \rangle}
\newcommand {\bkt} [1] {\langle #1 \rangle}
\newcommand {\dbkt} [2] {\langle #1 | #2 \rangle}
\newcommand {\tbkt} [3] {\langle #1 | #2 | #3 \rangle}
\newcommand {\pd} [2] {\frac{\partial #1}{\partial #2}}
\newcommand {\td} [2] {\frac{d #1}{d #2}}

 \newcommand {\beq}{\begin{equation}}
\newcommand {\eeq}{\end{equation}}
\newcommand {\bea}{\begin{eqnarray}}
\newcommand {\eea}{\end{eqnarray}}
\begin{document}
\title{Geometrical phase effects on the Wigner distribution of Bloch electrons}
\author{Dimitrie Culcer}
\author{Qian Niu}
\affiliation{Department of Physics, The University of Texas at
Austin, Austin TX 78712-1081.}
\date{\today}
\begin{abstract}
We investigate the dynamics of Bloch electrons using a density
operator method and connect this approach with previous theories
based on wave packets. We study non-interacting systems with
negligible disorder and strong spin-orbit interactions, which have
been at the forefront of recent research on spin-related
phenomena. We demonstrate that the requirement of gauge invariance
results in a shift in the position at which the Wigner function of
Bloch electrons is evaluated. The present formalism also yields
the correction to the carrier velocity arising from the Berry
phase. The gauge-dependent shift in carrier position and the Berry
phase correction to the carrier velocity naturally appear in the
charge and current density distributions. In the context of spin
transport we show that the spin velocity may be defined in such a
way as to enable spin dynamics to be treated on the same footing
as charge dynamics. Aside from the gauge-dependent position shift
we find additional, gauge-covariant multipole terms in the density
distributions of spin, spin current and spin torque.
\end{abstract}
 \maketitle

\section{Introduction}
Carrier dynamics in metals and semiconductors in the presence of
external electromagnetic fields, the potentials of which usually
vary on scales considerably large than the interatomic spacing,
have been conveniently described by semiclassical transport
theories. The semiclassical dynamics together with the Boltzmann
equation produce accurate descriptions of electrical and thermal
conduction\cite{AM}. In the larger picture, semiclassical
approaches are indispensable in problems involving both position
and momentum, since in quantum mechanics position and momentum
cannot be determined simultaneously. In recent years efforts have
been made to extend the semiclassical theory to spin transport and
generation\cite{Culcer, Bulkspin, Sinova}. The attempts to resolve
the challenges inherent in treating the transport of non-conserved
quantities constitute a vibrant ongoing effort\cite{Murakami I,
Murakami II, SZhang I, SZhang II, Hu, Inoue, Mishchenko,
Dimitrova, Yao I, Yao II}.

A fundamental feature of semiclassical transport is its accounting
for the finite extent of particles in real and reciprocal space.
This feature is most naturally incorporated into the dynamics of
wave packets\cite{Callaway, Ganesh} where the notion of a
wave-packet center in real space and $k$-space is retained. The
carrier dynamics are described in terms of the displacement of
these points under the action of external fields. The finite
extent of the wave packet has important consequences for transport
theory. For example, our recent research on spin transport has
shown that, due to the fact that the spin and charge centers of a
wave packet do not coincide, the expressions for the spin density,
torque and current distributions are, in the language of wave
packets, expressed as series of multipole terms\cite{Culcer}. It
has been demonstrated, in addition, that the wave packet formalism
captures the physics connected with adiabatic motion and the Berry
phase, in particular the Berry-curvature correction to the
semiclassical equations of motion\cite{Ganesh}. The Berry phase,
which until Berry's seminal article\cite{Berry} was not taken into
account, appears naturally as part of the wave packet distribution
function. The Berry-curvature correction to the wave packet
equations of motion is believed to play an important role in the
anomalous Hall effect\cite{AHE, Tomas, Yao} and in spin
transport\cite{Culcer, Bulkspin, Sinova, Murakami I, Murakami II},
among other phenomena.

Semiclassical transport theory is not restricted to wave packet
dynamics. Wave packets, which may be constructed out of one band
eigenstate\cite{Ganesh} or out of a superposition of eigenstates
of several degenerate bands\cite{Degenerate, Shindou}, represent
pure states. In order to treat mixed states (incoherent
superpositions of eigenstates) one must resort to a more general
formalism. This is the principal motivation behind the current
article.

The most general description of a quantum mechanical system is
based on the density operator. In this article we start from the
density operator and formulate a theory of carrier dynamics in
metals and semiconductors. We focus on non-interacting systems in
which disorder is weak, strong spin-orbit interactions are present
and a weak slowly-varying electric field is acting. These systems
have come under intense scrutiny in recent years along with the
take-off of spintronics\cite{Wolf, Hammar, Awschalom SC,
Wunderlich, Awschalom SG}. In such non-interacting systems the
formalism may be simplified by defining a reduced one-particle
density operator\cite{Reichl}. Since the Bloch bands are clearly
resolved the reduced density operator for this system may be
expanded in a basis of Bloch wave functions. The density matrix
which emerges from this expansion may be defined as a Wigner
function. This function is used to study single particle dynamics
and formulate definitions of macroscopic quantities.

We focus on a number of fundamental aspects of adiabatic particle
dynamics and demonstrate their relevance to transport phenomena.
We pay particular attention to the band mixing induced by the
electric field, which gives rise to a non-adiabatic correction to
the wave functions. We also address several important questions
concerning the relationship between the Wigner function formalism
and the wave-packet formalism\cite{Fiete}. We discuss the way the
carrier position is to be found and, where possible, compare the
result with the expression for the real-space center of a wave
packet. The requirement that the particle position be gauge
invariant results in a gauge-dependent shift in the position at
which the Wigner function is evaluated. This shift was also found
by Littlejohn and Flynn\cite{Littlejohn} in the study of
coupled-wave equations. This gauge-dependent shift is a
consequence of the freedom of choosing the real-space center of a
wave packet by changing the phase of the wave
function\cite{Ganesh, Marder}. The gauge-dependent shift in the
position of the Wigner function must be taken into account when
many-particle distributions, such as the particle number density,
are expressed in the crystal-momentum representation. The carrier
velocity may be derived directly from the particle position,
recovering the Berry-phase physics known from previous
work\cite{Ganesh}. We also show that a spin velocity may be
defined in such a way that spin transport may be described in an
analogous fashion to charge transport.

We discuss, in addition, important consequences of finite particle
size. The Wigner distribution is a quantum entity which takes into
account the finite extent of the particles in real and reciprocal
space. The distribution of, for example, spin for a single
carrier, may not coincide with that of charge and the macroscopic
spin distribution will be composed of a series of multipoles. The
spin current and spin torque distributions are in turn composed of
series of multipoles which we discuss and compare with those found
in our previous work on spin transport\cite{Culcer}.

The outline of this article is as follows. We present the
fundamentals of the single particle density matrix formalism in
section II. We determine the carrier position and velocity,
emphasizing the gauge-dependent shift in the former and the
Berry-phase correction in the latter. We calculate the particle
spin, torque and spin velocity and define a modified spin velocity
which satisfies an equation analogous to the charge velocity. In
section III we demonstrate the modifications which must be made to
extend the theory to many non-interacting particles. We define the
charge and current densities and show the equation of continuity
they satisfy. In the case of spin we define the spin, spin current
and spin torque distributions and show the equation of continuity
satisfied by them in the clean limit. In section IV we demonstrate
the effect of local gauge transformations and the modifications
which must be made to the dipoles in the charge- and spin-related
distributions in order to make them gauge covariant. We conclude
with a brief summary of our findings.

\section{Single-particle dynamics}
We consider systems described by a Hamiltonian having the
following general form:
\begin{equation}
\hat H = \hat H_0 + \hat H_{so} .
\end{equation}
The term $\hat H_0$ is composed of the usual kinetic-energy
contribution and the contribution due to the lattice-periodic
potential. The term $\hat H_{so}$ represents the spin-orbit
interaction term involving the carrier spins and the
lattice-periodic potential. We restrict our attention to the limit
in which this spin-orbit interaction is sizably stronger than the
disorder broadening and the thermal broadening. In this limit the
system may also be described in terms of well-defined bands. The
eigenstates of $\hat H$, which have the periodicity of the
crystal, are given by:
\begin{equation} \hat H \ket{\psi_m} =
\varepsilon_m\ket{\psi_m}.
\end{equation}
These wave functions are of the Bloch form, that is $|\psi_i
\rangle=e^{i{\bf k}\cdot{\bf \hat r}}|u_i\rangle $, where the
functions $\ket{u_i}$ represent the lattice-periodic parts of the
$\ket{\psi_i}$. The $\ket{u_i}$ are spinors with the full
periodicity of the lattice. Since the Hamiltonian contains strong
spin-orbit interaction terms, which may depend on wave vector and
position, it is not illuminating to decompose the eigenfunctions
into an orbital and a spin part.

\subsection{Density matrix}
We take the system under study to be described by a density
operator $\hat \rho$. It is not our concern in this article to
determine an expression for the density operator for a given
system, since methods of finding the density operator have been
studied extensively in the past\cite{Dyakonov I, Dyakonov II,
Dyakonov III, Dyakonov IV, Ivchenko, OptOrient, Aronov, QiZhang},
much work being done in the context of spin orientation of
carriers in the non-degenerate limit. We assume the form of $\hat
\rho$ to be known and study the role it plays in the dynamics of
charge and spin. The density operator may be expanded in the
Hilbert space spanned by a complete orthonormal set of Bloch wave
functions $\ket{\psi_{n{\bf k}}}$ as
\begin{equation}
\label{rhodef} \hat \rho = \sum_{n,n'}\sum_{{\bf k}, {\bf k}'}
\rho_{nn'}({\bf k}, {\bf k}') \ket{\psi_{n{\bf
k}}}\bra{\psi_{n'{\bf k}'}}.
\end{equation}
In the approach we follow, all the time dependence of the density
operator is contained in the wave functions, so that
$\rho_{nn'}({\bf k}, {\bf k}')$ does not have time dependence. In
thermal equilibrium it is diagonal and its elements,
$\rho_{nn}({\bf k}, {\bf k})$, are equal to the Fermi-Dirac
function $f_0(\varepsilon_n)$, where $\varepsilon_n$ is the band
energy.

The expectation value of any operator $\hat A$ is found from the
formula
\begin{equation}
\arraycolsep 0.3ex
\begin{array}{rl}
\bkt{\hat A} = & \displaystyle {\rm Tr}\, \hat\rho\hat A \equiv
\sum_{n,n'}\sum_{{\bf k}, {\bf k}'} \tbkt{\psi_{n{\bf k}}}{\hat
\rho}{\psi_{n'{\bf k}'}}\tbkt{\psi_{n'{\bf k}'}}{\hat
A}{\psi_{n{\bf k}}} \\ [2.2ex] = & \displaystyle {\rm tr} \,
\sum_{{\bf k}, {\bf k}'} \rho({\bf k}, {\bf k}') A({\bf k}', {\bf
k}).
\end{array}
\end{equation}
The notation $A({\bf k}', {\bf k})$ stands for the matrix elements
of the operator $\hat A$, namely ${A}_{n'n} = \tbkt{\psi_{n'{\bf
k}'}}{\hat A}{\psi_{n{\bf k}}}$. The operation denoted by tr is
simply the matrix trace, which does not include the wave-vector
summation. If the operator $\hat A$ is replaced by the identity
matrix we obtain ${\rm Tr}\, \hat \rho = \sum_{{\bf k}} {\rm tr}\,
\rho({\bf k}, {\bf k})$. For a single particle, the normalization
condition on the density operator is ${\rm Tr}\, \hat\rho = 1$.

In order to make transparent the analogy with the language of wave
packets, center of mass and relative coordinates may be defined in
$k$-space such that {\bf Q} = ${\bf k} - {\bf k}'$ and ${\bf q} =
\frac{{\bf k} + {\bf k}'}{2}$. The density operator as a function
of these coordinates can be re-expressed as
\begin{equation}
\hat\rho = \sum_{n, n'}\sum_{{\bf q}, {\bf Q}} \rho_{nn'}( {\bf
q}_+, {\bf q}_-) \ket{\psi_{n+}}\bra{\psi_{n'-}},
\end{equation}
where $\ket{\psi_{n\pm}} = e^{i{\bf q}_\pm \cdot \hat{\bf
r}}\ket{u_{n\pm}}$ and $\ket{u_{n\pm}}$ is the periodic part of
the Bloch wave at ${\bf q}_\pm = {\bf q} \pm \frac{{\bf Q}}{2}$.

The Wigner function corresponding to the one particle density
matrix is found through the transformation
\begin{equation}
\label{wigout} \rho_{nn'}({\bf q}_+, {\bf q}_-) = \int d^3r\,
e^{-i{\bf Q}\cdot {\bf r}}\rho_{nn'}({\bf q}, {\bf r}).
\end{equation}
For the sake of concreteness the integrals are represented as
three dimensional. Nevertheless, the theory applies to systems of
any dimensionality. The Wigner function plays the role of a
distribution function in the variables {\bf q} and {\bf r}.
Technically, however, it is not a distribution since it may take
negative values\cite{Reichl}. The inverse transformation is
\begin{equation}
\label{wigin} \rho_{nn'}({\bf q}, {\bf r}) = \int
\frac{d^3Q}{(2\pi)^3}\, e^{i{\bf Q}\cdot{\bf r}}\rho_{nn'} ({\bf
q}_+, {\bf q}_-).
\end{equation}
Finally, replacing the vector summations by integrations we are
able to represent the density operator in the following form
\begin{widetext}
\begin{equation}
\begin{split}
\hat\rho = \sum_{nn'}\int\,d{\cal V}\, \rho_{nn'}({\bf q},
{\bf r}) \int d^3Q\, e^{i{\bf q}_+ \cdot(\hat{\bf
r} - {\bf r})} \ket{u_{n+}} \bra{u_{n'-}} e^{-i{\bf q}_-
\cdot(\hat{\bf r} - {\bf r})},
\end{split}
\end{equation}
\end{widetext}
where $\int\,d{\cal V} = \int\!\int \frac{d^3q\, d^3r\,}{(2\pi)^3}$.
For the remainder of this section, the Wigner function $\rho_{nn'}({\bf q}, {\bf r})$ will
frequently be abbreviated to $\rho$. The variables {\bf q} and {\bf r} are
simply labels for the carriers, not physical observables.
In particular, the dummy variable {\bf r} in
the Fourier expansion of the density matrix is
simply the Fourier dual of {\bf Q} and must not be
confused with the position operator $\hat{\bf r}$ appearing in the
density operator. It does not correspond to an actual position.

\subsection{Carrier position}
If we consider a particle in one band, labelled $n$, $\rho$
reduces to a scalar, which will be denoted by $\rho_n$. The
position of the particle can be found as the expectation value of
the position operator, which yields
\begin{equation}
\label{r} {\rm Tr} \, \hat\rho\hat{\bf r} = \int\,d{\cal V}\,
\rho_n({\bf q}, {\bf r}) ({\bf r} + {\cal R}_n).
\end{equation}
Here ${\cal R}_n \equiv {\cal R}_{nn} = \dbkt{u_n}{i\pd {u_n}{{\bf
q}}}$. The integrand is not gauge invariant but the integral can
be shown to be by changing the variable of integration {\bf r} to
${\bf r}' = {\bf r} + {\cal R}_n$. The connection ${\cal R}_n$ has
no position dependence, therefore the Jacobian of the
transformation is unity. The expectation value of the position
operator is
\begin{equation}\label{rnew}
{\rm Tr}\, \hat\rho\hat{\bf r} = \int\,d{\cal V}'\, \rho_n({\bf q}, {\bf r}' - {\cal
R}_n)\, {\bf r}',
\end{equation}
where $d{\cal V}'$ is defined in the same way as $d{\cal V}$ but
with {\bf r} replaced by ${\bf r}'$. The gauge invariance of
(\ref{rnew}) will emerge below. We conclude that ${\bf r}$ is to
be interpreted as a label for the charge carrier, while the
effective particle position is ${\bf r}' = {\bf r} + {\cal R}_n$.
Neither the label ${\bf r}$ nor the gauge field ${\cal R}$ are by
themselves gauge invariant, but together they form a
gauge-invariant quantity which represents the true position of the
carrier. This result was also found earlier, in somewhat different
circumstances, in the work of Littlejohn and
Flynn\cite{Littlejohn}. In the one-band limit, a clear connection
can also be made with the dynamics of wave packets. The
gauge-dependent shift in {\bf r} reflects the freedom of changing
the phase of the wave functions $\ket{\psi_{n{\bf k}}}$,
$\ket{\psi_{n'{\bf k}'}}$ in (\ref{rhodef}). It is the same
freedom one has in defining the center of mass of a wave packet,
demonstrated by Sundaram and Niu\cite{Ganesh}, by changing the
overall phase of the wave function\cite{Marder}.

For multiple bands, the expression for the particle position is
(the Einstein summation convention will be used henceforth)
\begin{equation}
\label{rmb} {\rm Tr}\, \hat\rho\hat{\bf r} = \int\,d{\cal V}\, \rho_{nn'}({\bf q}, {\bf r}) ({\bf
r}\delta_{n'n} + {\cal R}_{n'n}).
\end{equation}
One can rewrite this expression as
\begin{equation}
{\rm Tr}\, \hat\rho\hat{\bf r} = {\rm tr}\int\,d{\cal V}\, \rho({\bf q}, {\bf r}) \, ({\bf r} +
{\cal R}).
\end{equation}
and make the substitution ${\bf r}' = {\bf r} + {\cal R}$ as in
the single band case. Although ${\cal R}$ is a matrix the Jacobian
of this transformation is unity. Finally, the expectation value of
the position operator can be expressed formally as
\begin{equation}
\label{rmbnew} {\rm Tr}\, \hat\rho\hat{\bf r} = {\rm tr}\int\,d{\cal V}'\, \rho({\bf q}, {\bf r}' - {\cal
R})\, {\bf r}'.
\end{equation}
Unlike the single band case, the expression $\rho({\bf q}, {\bf
r}' - {\cal R})$ is a formal abbreviation for the Taylor expansion
about ${\bf r}'$, that is $\rho({\bf q}, {\bf r}') - {\cal
R}\cdot\nabla_{{\bf r}'}\rho({\bf q}, {\bf r}')$ $+$ higher order.

Expression (\ref{rnew}) identifies the position of a particle. To
determine if a particle is localized at its position one must
calculate the variance of the position operator, the expectation
value $\bkt{\hat {\bf r}^2} - \bkt{\hat{\bf r}}^2$, and ensure
that is does not diverge. It is shown in the appendix that the
variance does not diverge and that this result applies to any
number of bands.

\subsection{Carrier velocity in an electric field}
We consider a system acted on by a weak external electric field.
The effect of this electric field is incorporated fully into the
gauge invariant crystal momentum through the addition of the
electromagnetic vector potential ${\bf A} = -{\bf E}t$. Because
the Bloch functions retain translational symmetry the
electromagnetic vector potential does not enter the
travelling-wave part of the wave functions, which have the form
$\ket{\psi_{n\tilde{\bf q}}} = e^{i{\bf q}\cdot\hat{\bf
r}}\ket{u_{n\tilde{\bf q}}}$. The lattice-periodic functions
$\ket{u_{n\tilde{\bf q}}}$ depend implicitly on time only through
the crystal wave vector $\tilde{\bf q} = {\bf q} + \frac{e{\bf
A}}{\hbar}$. However, the presence of the external electric field
results in a non-adiabatic mixing of the bands with the result
that the perturbed lattice-periodic functions, $\ket{\bar
u_{m\tilde{\bf q}}}$, have the following form to first order in
the electric field\cite{Messiah}
\begin{equation}
\label{nonadb} \ket{\bar u_{m\tilde{\bf q}}} =
e^{-i\phi_m}\left(\ket{u_{m\tilde{\bf q}}} - \sum_{n\ne
m}\frac{\tbkt{u_{n \tilde{\bf q}}}{i\hbar\td{}{t}}{u_{m\tilde{\bf
q}}}}{\varepsilon_m - \varepsilon_n}\ket{u_{n\tilde{\bf
q}}}\right).
\end{equation}
The phase $\phi_m(\tilde{\bf q}, t)$ includes the dynamical phase
and the Berry phase. The differential $\td{}{t}$ is equivalent to
$\dot{\tilde{\bf q}}\cdot\pd{}{\tilde{\bf q}}$ since
$\pd{}{t}\ket{u_{n\tilde{\bf q}}} = 0$. The result expressed by
(\ref{nonadb}) is general. Moreover, although its derivation
relies on the assumption that the bands are non-degenerate it can
be shown that, when calculating {\it intrinsic} contributions to
transport (for the definition of intrinsic please see the appendix
and our recent work\cite{Bulkspin}), the result also holds for
degenerate bands with the difference that the sum must exclude
{\it all} bands which are degenerate in energy with band $m$. The
proof of this statement is given in the appendix. Henceforth,
$\ket{u_{m\tilde{\bf q}}}$ and $\ket{\bar u_{m\tilde{\bf q}}}$
will be abbreviated to $\ket{u_{m}}$ and $\ket{\bar u_{m}}$
respectively. As stated above, given that the $\ket{u_m}$ are
functions of $\tilde{\bf q}$ only, one may replace $\td{}{t}$ in
(\ref{nonadb}) by $\dot{\tilde{\bf q}}\cdot\pd{}{\tilde{\bf q}}$,
where $\dot{\tilde{\bf q}} = $$-\frac{e{\bf E}}{\hbar}$. Equation
(\ref{nonadb}) can then be written as
\begin{equation}
\label{barufin} \ket{\bar u_m} = e^{-i\phi_m}\left(\ket{u_m} +
e{\bf E}\cdot\sum_{n\ne m}\frac{{\cal R}_{nm}} {\varepsilon_m -
\varepsilon_n}\ket{u_n}\right),
\end{equation}
where the connection ${\cal R}_{nm} = \tbkt{u_n}{i\pd{}{\tilde{\bf
q}}}{u_m}$. The $\ket{\bar u_m}$ form a complete set. They are,
however, {\it not} eigenstates of the {\it time-dependent}
Hamiltonian $\tilde H \equiv \tilde H (\tilde{\bf q})$.

In the evaluations of matrix elements in this paper
the only property of the basis functions that is used is their
Bloch periodicity. Therefore the results which are expressed in
terms of the lattice periodic Bloch functions hold as well for the
$\ket{\bar u_n}$ as for the $\ket{u_n}$.

In the absence of disorder, since all the time dependence is
contained in the wave functions, the density matrix
$\rho_{nn'}({\bf k}, {\bf k}')$ in (\ref{rhodef}) can only depend
on the wave vector {\bf k}, not on the crystal wave vector
$\tilde{\bf k}$. As a result the Wigner function $\rho_{nn'}({\bf
q}, {\bf r})$ only depends on the wave vector {\bf q}, not on
$\tilde{\bf q}$.

It is customary to consider only a subset of the Hilbert space
which contains the bands that are relevant to transport, which in
semiconductors usually refers to the topmost filled valence bands
and/or the lowest filled conduction bands. Since the gauge of the
$\ket{\bar u_m}$ is not fixed we impose the following gauge-fixing
condition {\it in the subspace} under consideration
\begin{equation}\label{gf}\arraycolsep 0.3ex
\begin{array}{rl}
\bra{\bar u_n}i\hbar\td{}{t}{\ket{\bar u_m}} = & \displaystyle
\bra{\bar u_n} {\tilde{\mathcal {H}}} \ket{\bar u_m}, \,\,\, {\rm
for} \,\,\, n \,\,\, {\rm in \,\,\, subspace} \\ [2.2ex]
 & \displaystyle 0, \,\,\, {\rm for} \,\,\, n \,\,\, {\rm out \,\,\, of \,\,\,
subspace},
\end{array}\end{equation}
where $\tilde{\mathcal{H}} = e^{-i{\bf q}\cdot\hat{\bf r}}{\tilde
H}e^{i{\bf q}\cdot\hat{\bf r}} $. This condition fixes the
phase(s) $\phi_m$ in (\ref{nonadb}). We shall henceforth work only
with the basis set $\{ \ket{\bar u_n} \}$.

The particle velocity can be derived directly from (\ref{rmb}) by
evaluating the time derivative
\begin{equation}\label{drdt}
\td{}{t}\, {\rm Tr}\,\hat\rho\hat{\bf r} = {\rm tr} \int d{\cal V}\,
\rho \, \td{\bar{\cal
R}}{t}.
\end{equation}
Using equation (\ref{gf}), the differential $\td{\bar{\cal R}}{t}$
becomes, after a little straightforward algebra,
\begin{equation}\begin{split}\label{dRdt}
\td{\bar{\cal R}_{n'n}}{t} = \frac{1}{\hbar} \pd{\bar
H_{n'n}}{\tilde{\bf q}} + i \left( \dbkt{\td{\bar u_{n'}}{t}}{\pd{\bar
u_n}{\tilde{\bf q}}} - \dbkt{\pd{\bar u_{n'}}{\tilde{\bf
q}}}{\td{\bar u_n}{t}}\right).
\end{split}\end{equation}
The abbreviation $\bar H_{n'n}$ stands for the matrix elements of
the time-dependent Hamiltonian $\tilde{H}$ in the presence of an
electric field, $\tbkt{\bar u_{n'}}{\tilde H}{\bar u_n}$. In one
band it is easily shown that (\ref{dRdt}) becomes
\begin{equation}
\label{vel} \td{\bar{\cal R}_{n}}{t} = \frac{1}{\hbar}
\pd{\varepsilon_n}{\tilde{\bf q}} - \dot{\tilde{\bf q}} \times
\bar{\bf \Omega}_{\it n} + \bar{\bf \Xi}_{\it n}^{{\it t}{\bf q}}.
\end{equation}
The Berry curvature for band $n$, $\bar{\bf \Omega}_n$, is given
by $\nabla_{\tilde{\bf q}} \times \bar{\cal R}_n$ and
$\nabla_{\tilde{\bf q}} \times$ represents the curl operator in
reciprocal space. The quantity $\bar{\bf \Xi}^{t{\bf q}}_{\it n}$
is defined as $i \left( \dbkt{\pd{\bar u_{n'}}{t}}{\pd{\bar
u_n}{\tilde{\bf q}}} - \dbkt{\pd{\bar u_{n'}}{\tilde{\bf
q}}}{\pd{\bar u_n}{t}}\right)$. Equation (\ref{vel}) is the analog
of the semiclassical equation of motion found by Sundaram and
Niu\cite{Ganesh}. Writing $\dot{\tilde{\bf q}} =$$-\frac{e{\bf
E}}{\hbar}$ we obtain for electrons in a single band:
\begin{equation}
\label{rhovfin}
\begin{split}
\td{}{t}\, {\rm Tr}\,\hat\rho\hat{\bf r} = \int d{\cal V}\, \rho_n
\left(\frac{1}{\hbar} \pd{\varepsilon_n}{\tilde{\bf q}} + \frac{e
{\bf E}}{\hbar} \times \bar{\bf \Omega}_n  + \bar{\bf \Xi}^{t{\bf
q}}_{\it n}\right).
\end{split}
\end{equation}
For multiple bands, an elegant result is obtained by introducing
the covariant derivatives\cite{Degenerate} $\frac{D}{D\tilde{\bf
q}} = \pd{}{\tilde{\bf q}} - i{\cal R}$ and $\frac{D}{Dt} =
\td{}{t} - i{\cal A}$, where ${\cal A}_{n'n} =
\dbkt{u_{n'}}{i\td{u_n}{t}}$. Making use of these derivatives,
equation (\ref{dRdt}) can be written in the manifestly gauge
covariant form
\begin{equation}\begin{split}\label{dRdtcov}
\td{\bar{\cal R}^\alpha_{n'n}}{t} =
\frac{1}{\hbar}[\frac{D}{D\tilde{q}_\alpha}, \bar H]_{n'n} -
i[\frac{D}{D\tilde{q}_\alpha}, \frac{D}{Dt}]_{n'n}.
\end{split}\end{equation}
It is important to point out that, from the gauge-fixing condition
(\ref{gf}) which defines ${\cal A}$, it is evident that ${\cal A}$
is gauge covariant, implying that the time derivative $\td{}{t}$
is itself gauge-covariant. This is a peculiarity of the
gauge-fixing condition we have chosen.

\begin{figure}[tbp]
\centering \epsfig{file=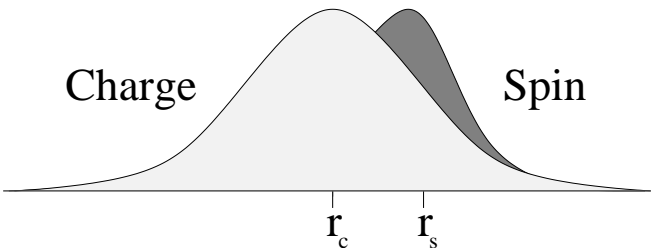, width=\columnwidth}
\caption{For a particle of finite extent the charge and spin
distributions in real space are in general do not coincide. The
same is true of the charge and spin distributions in reciprocal
space.} \label{fig:wavepacket}
\end{figure}

\subsection{Carrier spin, spin torque and consequences of finite particle size}

The fact that carriers have a finite extent in real and
wave-vector space has profound implications for particle dynamics
and transport. These implications were pointed out in a previous
publication \cite{Culcer} using the semiclassical language of
wave-packets and will be elaborated in this section from the
density matrix point of view. For the sake of clarity we will
specialize in spin, although the discussion in this section
applies to any quantity.

When considering the transport of spin in a system composed of
many carriers, one must associate a spin distribution with each
individual carrier. A center of spin may be defined for each
particle by ${\bf r}_{s} = \frac{\bkt{\hat{\bf r}\hat
s}}{\bkt{\hat s}}$. This definition may be troublesome if the
expectation value of the spin operator were zero, but its only
purpose is to illustrate a physical principle. Evidently, if one
replaced the spin operator with the charge or mass one would
obtain the particle position as given by (\ref{r}), which we will
refer to as ${\bf r}_c$. It is obvious from the definition of
${\bf r}_s$ that in the general case there is no reason for the
center of the spin distribution of an individual particle to be
the same as the actual particle position. This center will be
different for each component of the spin operator. Furthermore,
since spin is not conserved in the presence of e.g. spin-orbit
interactions, ${\bf r}_s$ may also be a function of time. This
suggests that the spin distribution of one carrier will in general
have a different shape than its charge distribution, and that the
time development of the two distributions may be quite different.
These facts are illustrated in Figs.~1 and 2.

To evaluate the average carrier spin consider the expectation
value of the operator $\hat s$, which stands for any one component
of the spin operator. The result is
\begin{equation}
\begin{split}
{\rm Tr}\,\hat\rho\hat s = {\rm tr} \int\,d{\cal V}\, \rho\, \bar
s,
\end{split}
\end{equation}
where $\bar s_{n'n} = \tbkt{\bar u_{n'}}{\hat s}{\bar u_n}$ are
the matrix elements of the spin operator. In the presence of
spin-orbit interactions a spin torque is associated with the spin
of every carrier, which accounts for the non-conservation of spin,
as shown in Fig.~2. The average carrier torque is found by
evaluating the expectation value of $\hat\tau =
\frac{i}{\hbar}[\tilde H, \hat s]$. The result is
\begin{equation}
\begin{split}
{\rm Tr}\,\hat\rho\hat \tau = {\rm tr} \int\,d{\cal V}\, \rho\,
\bar \tau,
\end{split}
\end{equation}
where $\bar \tau_{n'n} = \tbkt{\bar u_{n'}}{\hat \tau}{\bar u_n}$.

As a result of the distinction between ${\bf r}_s$ and ${\bf
r}_c$, in calculations of spin-related quantities which use the
center of charge as the reference, multipole terms must be taken
into account in addition to the average quantities calculated
above. For example, in the distribution of spin a spin dipole will
be present, as well as higher order multipole terms, which are
assumed small. The spin dipole is found as the expectation value
$\bkt{\hat{\bf r}\hat s}$, evaluated in the appendix, and yields
the gauge field
\begin{equation} \bar{\bf M}^s_{n'n} =
\frac{i}{2} (\tbkt{\bar u_{n'}}{\hat {s}}{\pd{\bar u_n}{{\bf q}}}
- \tbkt{\pd{\bar u_{n'}}{{\bf q}}}{\hat {s}}{\bar u_{n}}).
\end{equation}
In contrast to ${\bf r}_s$, $\bar{\bf M}^s$ is well defined.
Similarly, in the distribution of torque of an individual particle
a torque dipole will be present. The torque gauge field $\bar{\bf
M}^\tau = \bkt{\hat{\bf r}\hat\tau}$ is given, as shown in the
appendix, by
\begin{equation}
\bar{\bf M}^\tau_{n'n} = \frac{i}{2} (\tbkt{\bar u_{n'}}{\hat
{\tau}}{\pd{\bar u_n}{{\bf q}}} - \tbkt{\pd{\bar u_{n'}}{{\bf
q}}}{\hat {\tau}}{\bar u_{n}}).
\end{equation}
$\bar{\bf M}^s$ and $\bar{\bf M}^\tau$ are gauge dependent. It
will prove useful to define gauge-covariant dipoles as $\bar{\bf
p}^s = \bar{\bf M}^s - \frac{1}{2}\{ \bar{\cal R}, \bar s \}$ and
$\bar{\bf p}^\tau = \bar{\bf M}^\tau - \frac{1}{2}\{ \bar{\cal R},
\bar\tau \}$ respectively. The remainder of the discussion of
these gauge-covariant spin and torque dipoles is deferred to
section IV.
\begin{figure}[tbp]
\centering \epsfig{file=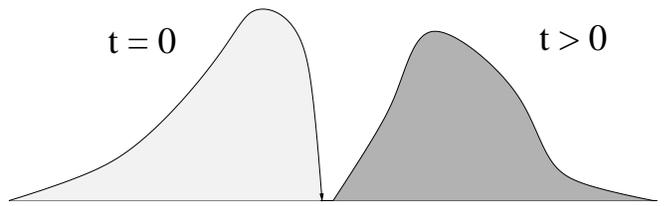, width=\columnwidth}
\caption{In the presence of spin-orbit interactions the spin
distribution of a particle changes in time. The horizontal axis
may represent position or wave vector.} \label{fig:spinintime}
\end{figure}

\subsection{Carrier spin velocity in an electric field}
We define the spin velocity as the expectation value $\bkt{\hat
s\hat{\bf v}}$, where products of non-commuting operators are
assumed to be symmetrized. The result is
\begin{equation}
\begin{split}
{\rm Tr}\,\hat\rho\hat s\hat{\bf v} = {\rm tr} \int\,d{\cal V}\,
 \rho \, \bar{\bf v}^s.
\end{split}
\end{equation}
The spin velocity $\bar{\bf v}^s$ is given by
\begin{equation}
\arraycolsep 0.3ex
\begin{array}{rl}
\bar{\bf v}^s = &\displaystyle \td{}{t}\bkt{\hat{\bf r}\hat{s}} -
\bkt{\hat{\bf r}\hat{\tau}} \\ [2.2ex] = &\displaystyle
\td{\bar{\bf M}^s}{t} - \bar{\bf M}^\tau.
\end{array}
\end{equation}
In order to write the velocity in a gauge covariant form, we
replace in the above $\bar{\bf M}^s = \bar{\bf p}^s +
\frac{1}{2}\{ \bar{\cal R}, \bar s \}$ and $\bar{\bf M}^\tau =
\bar{\bf p}^\tau + \frac{1}{2}\{ \bar{\cal R}, \bar \tau \}$. This
yields immediately
\begin{equation}\begin{split}\label{vsps}
\bar{\bf v}^s = \frac{1}{2}\{ \td{\bar{\cal R}}{t}, \bar s\} +
\td{\bar{\bf p}^s}{t} - \bar{\bf p}^\tau .
\end{split}\end{equation}
Equation (\ref{vsps}) is the gauge-covariant form of equation (10)
in Ref. [2], the integrand of which represents the spin velocity.
The first term in (\ref{vsps})is a convective contribution and
represents a moving electron transporting its average spin along
with it. The second term is the time derivative of the spin dipole
while the last term is the torque dipole which takes into account
the non-conservation of spin. We may incorporate $\bar{\bf
p}^\tau$ into a modified spin velocity\cite{Culcer, Junren}, which
we shall call $\bar{\bf v}^t$, by $\bar{\bf v}^t = \bar{\bf v}^s +
\bar{\bf p}^\tau$. The modified spin velocity is given simply by:
\begin{equation}\begin{split}\label{vtps}
\bar{\bf v}^t = \frac{1}{2}\{ \td{\bar{\cal R}}{t}, \bar s\} +
\td{\bar{\bf p}^s}{t}.
\end{split}\end{equation}
The importance of this definition of the velocity will be seen in
the following section, when the macroscopic spin current is
introduced.

\section{Many particle distributions}
We will consider the macroscopic distributions of charge and spin
starting from the formalism we have developed. In particular, our
discussion of the corrections to the spin density, current and
torque distributions is motivated by the observation that the
existing literature has omitted various contributions to spin
transport. Several works have used semiclassical concepts but did
not arrive at answers containing all the terms we have derived.

The particle number density is defined by the formula:
\begin{equation}
\label{n} n ({\bf R}, t) = {\rm Tr} [\hat \rho \delta( {\bf R} -
\hat{\bf r} ) ].
\end{equation}
The trace operation Tr involves integration over {\bf r}. The
procedure we follow has affinities with the coarse graining of
electrodynamics in material media \cite{Jackson}. The size of the
carriers is taken to be smaller than the length scale of the
Wigner function. We regard the $\delta$-function as a sampling
function\cite{Jackson} with a width somewhere between the
microscopic scale of the carriers and the macroscopic scale of the
Wigner function. The rationale for regarding the $\delta$-function
as a sampling function comes from the realization that often the
physics of the problem does not require absolute resolution of
position, provided that the resolution is finer than the length
scale of the external field. Since the variance of $\hat{\bf r}$
is finite, the $\delta$-function may be expanded in a Taylor
series about the dummy variable {\bf r} in the form $ \delta( {\bf
R} - \hat{\bf r} ) = \delta( {\bf R} - {\bf r} )  - \nabla_{\bf R}
\cdot(\hat{\bf r} - {\bf r}) \delta( {\bf R} - {\bf r} ) + {\cal
O} [(\hat{\bf r} - {\bf r})^2] $. The expansion is truncated at
the first order for simplicity but we will show in the next
section that all the terms may be recovered in a concise and
elegant fashion. The number density can be expressed as:
\begin{equation}\label{ndip}
n ({\bf R}, t) = {\rm tr} \int d^3 q\, (\rho -
\nabla_{\bf R}\cdot \rho \,\bar{\mathcal R}).
\end{equation}
In the above $\rho$ stands for $\rho({\bf q}, {\bf R})$, an abbreviation
which will be frequently used in the remainder of the article. Note
that the gauge field $\bar{\cal R}$ in (\ref{ndip}) plays the role of a dipole.

\subsection{Electrical charge and current densities}

The charge density is defined by the formula
\begin{equation}
n^q({\bf R}, t) = q\, n ({\bf R}, t) = q\, {\rm Tr} [\hat \rho
\delta( {\bf R} - \hat{\bf r} ) ].
\end{equation}
The charge $q$ is not to be confused with the wave vector {\bf q}.
The charge current density ${\bf J}^q$ is defined by the equation
\begin{equation}
{\bf J}^q({\bf R}, t) = q\, {\rm Tr} [\hat \rho \delta( {\bf R} -
\hat{\bf r} ) \hat{\bf v}].
\end{equation}
The charge equation of continuity in the absence of external
sources and disorder,
\begin{equation}
\pd{n^q}{t} + \nabla\cdot{\bf J}^q = 0,
\end{equation}
is readily verified from the first-principles definitions of the
charge and current densities.

When the $\delta$-function is expanded the current density takes
the form
\begin{equation}\label{jq}
{\bf J}^q({\bf R}, t) = q\,{\rm tr}  \int \frac{d^3q}{(2\pi)^3} \,
\rho \, \bar{\bf v}.
\end{equation}
The velocity matrix elements $\bar{\bf v}_{n'n} = \tbkt{\bar
u_{n'}}{\tilde{\bf v}}{\bar u_{n}}$ as shown in the appendix,
where $\tilde{\bf v} = e^{-i{\bf q}\cdot\hat{\bf r}}\hat{\bf
v}e^{i{\bf q}\cdot\hat{\bf r}}$. Since in the equation of
continuity it is the gradient of the current that appears, we have
not included in (\ref{jq}) corrections to the current which arise
from the fact that the velocity distribution of a single carrier
is different from its charge distribution. These corrections are
in principle present but have been omitted for simplicity.

In homogeneous systems the gradient term in (\ref{jq}) drops out
of the equation of continuity and, based on equations (\ref{drdt})
and (\ref{jq}), the charge current can be expressed as the time
rate of change of the gauge field $\bar{\cal R}$
\begin{equation}\label{jqr}
{\bf J}^q({\bf R}, t) = q\, {\rm tr} \int \frac{d^3q}{(2\pi)^3} \,
\rho\,\td{\bar {\cal R}}{t}.
\end{equation}

\subsection{Spin, spin current and spin torque densities}
The spin density is defined as
\begin{equation}
S({\bf R}, t) = {\rm Tr} [\hat \rho \delta( {\bf R} - \hat{\bf r}
) \hat s ],
\end{equation}
while the torque density takes the form
\begin{equation}
{\cal T}({\bf R}, t) = {\rm Tr} [\hat \rho \delta( {\bf R} -
\hat{\bf r} ) \hat \tau ]
\end{equation}
and the spin current density is defined as
\begin{equation}
{\bf J}^s({\bf R}, t) = {\rm Tr} [\hat \rho \delta( {\bf R} -
\hat{\bf r} )\hat s \hat{\bf v}].
\end{equation}
In the absence of disorder the spin equation of continuity is
\begin{equation}
\pd{S}{t} + \nabla\cdot{\bf J}^s = {\cal T},
\end{equation}
which is verified from the first-principles definitions introduced
above.

The ${\delta}$-functions are expanded in the same way as for the
particle number density, whereupon the spin density can be
expressed as
\begin{equation}
\label{s} S({\bf R}, t) = {\rm tr} \int \frac{d^3q}{(2\pi)^3} \,
(\rho\, \bar s - \nabla_{\bf R} \cdot \rho\,\bar{\bf M}^s),
\end{equation}
and the torque density is
\begin{equation}
{\cal T}({\bf R}, t) = {\rm tr} \int \frac{d^3q}{(2\pi)^3} \,
(\rho\, \bar\tau - \nabla_{\bf R} \cdot \rho\,\bar{\bf M}^\tau).
\end{equation}
Ignoring the gradient term, the spin current is
\begin{equation}
{\bf J}^s({\bf R}, t) = {\rm tr} \int \frac{d^3q}{(2\pi)^3}
\,\rho\,\bar{\bf v}^s.
\end{equation}
In analogy with the modified spin velocity of the previous section, a modified spin current ${\bf J}^t$ may be defined by:
\begin{equation}
\arraycolsep 0.3ex
\begin{array}{rl}
{\bf J}^t({\bf R}, t) = & \displaystyle {\rm tr} \int
\frac{d^3q}{(2\pi)^3} \,\rho\,\bar{\bf v}^t \\ [2.2ex] = &
\displaystyle {\rm tr} \int \frac{d^3q}{(2\pi)^3}
\,\rho\,\left(\frac{1}{2}\{\td{\bar{\cal R}}{t}, \bar s\} +
\td{\bar{\bf p}^s}{t}\right).
\end{array}\end{equation}
The equation of continuity satisfied by this current in the clean limit is:
\begin{equation}
\pd{S}{t} + \nabla\cdot{\bf J}^t = {\rm tr} \int
\frac{d^3q}{(2\pi)^3} \,\rho\,\bar\tau .
\end{equation}
In many models, such as the spherical four-band Luttinger
model\cite{Luttinger}, the RHS vanishes when all bands in the
subspace are taken into account. In that case the spin current
${\bf J}^t$ is conserved. Much effort has been devoted to finding
a conserved spin current. In our previous work \cite{Culcer,
Junren} we have argued, based on semiclassical ideas, that the
closest one can come to a conserved spin current is by including
the torque dipole in the definition of the current. In this
article we have derived this current from a density matrix point
of view and show that it supports our earlier conclusions. To
date, the definition ${\bf J}^t$ is the closest one has come to a
conserved spin current. Moreover, as shown also by Zarea and Ulloa
\cite{Zarea}, the behavior of the modified spin current may be
rather different from that of ${\bf J}^s$.

\section{Gauge transformations}

We have argued in section II that in discussing transport of an
observable one must take into account the fact that the
distribution of that observable for each individual carrier in
general involves a dipole correction, and in principle also
higher-order corrections. If the quantity being transported is not
conserved then the determination of its rate of change also
involves a dipole correction. The existence of what appears to be
a sound physical argument for the inclusion of these corrections
suggests that objects such as the spin dipole and the torque
dipole are not simply mathematical artifacts meant to facilitate
calculations. It is appropriate to investigate whether they are in
fact fundamental quantities with a true physical meaning. If that
were the case, they ought to be expressible in a gauge-covariant
way and to give rise to observable effects.

In addition to these considerations, our aim of providing a
physically transparent formalism requires a test of the gauge
covariance of the macroscopic quantities and equations of motion
we have derived in this article. The physics contained in them
must be independent of the choice of basis functions. Furthermore,
the physical discussion can become cluttered with futile and
meaningless objects if it is formulated in terms of
gauge-dependent objects. In contrast, the gauge-covariant
expressions we derive below are simple, elegant and transparent.

A general local gauge transformation is represented by $ \ket{\bar
u_{m}} \rightarrow O_{mn}\ket{\bar u_{n}} $, with $ O_{mn} =
O_{mn}(\tilde{\bf q})$ and $m$ and $n$ are indices of bands in the
subspace under consideration. Under this operation
\begin{equation}\label{Rtrans}
\bar{\cal R} \rightarrow \tilde{\cal R} + i(O^{-1}
\pd{O}{\tilde{\bf q}}),
\end{equation}
in which $\tilde{\cal R}$ stands for $O\bar{\cal R}O^{-1}$. The
Berry curvature for one band $\bar{\bf \Omega}_n$ is discussed
extensively in the paper of Sundaram and Niu\cite{Ganesh}. One
important additional detail which emerges from the transformation
given by (\ref{Rtrans}) and the definition $\bar{\bf \Omega}_n =
\nabla_{\tilde{\bf q}} \times \bar{\cal R}_n$ is the fact that, if
the curvature is non-zero, one cannot make a gauge transformation
to eliminate $\bar{\cal R}_n$. It can be easily shown that,
whereas $\bar{\cal R}_n$ is gauge dependent, the Berry curvature
is gauge covariant. Therefore if $\bar{\bf \Omega}_n$ is non-zero
in one gauge it is non-zero in all gauges and there is no gauge in
which $\bar{\cal R}_n$ can be zero. As a result, in systems in
which the curvature is non-zero the gauge-dependent position shift
is necessarily present in the Wigner function. It can be regarded
as a `penalty' for working with the position operator in a basis
of definite wave vector.

\subsection{Gauge covariance of macroscopic densities}

The Wigner function itself changes under the gauge transformation
defined above. Using (\ref{wigout}), we find that, to first order in $O$, the
Wigner function changes as
\begin{equation}
\label{rhogt} \rho \rightarrow \tilde\rho + \frac{i}{2}
\nabla_{\bf R} \cdot \{ \tilde\rho, O^{-1} \pd{O}{\tilde{\bf q}}
\},
\end{equation}
where $\tilde\rho = O^{-1}\rho O$ (the opposite of the gauge
connection $\bar{\cal R}$).

All the macroscopic densities defined in the previous sections are
covariant under a local gauge transformation. This will be
demonstrated for the spin density. Under the above gauge
transformation, the gauge field $\bar{\bf M}^s$ transforms as
\begin{equation}\begin{split}
\bar{\bf M}^s \rightarrow  \tilde{\bf M}^s +
\frac{i}{2}\{ {\tilde s}, O^{-1}\pd{O}{\tilde{\bf q}} \}.
\end{split}\end{equation}
The matrix elements of the transformed gauge field $\tilde{\bf
M}^s$ are given by $\tilde{\bf M}^s_{n'n} = \frac{i}{2}(
\tbkt{\bar u_{n'}}{ \tilde{s}}{\pd{\bar u_n}{\tilde{\bf q}}} -
\tbkt{\pd{\bar u_{n'}}{\tilde{\bf q}}}{\tilde{s}}{\bar u_{n}}) $
and the transformed spin operator $\tilde s = O^{-1}\hat s O$. To
first order the extra terms acquired by the spin density under a
gauge transformation are
\begin{equation}
{\rm tr} ( \tilde\rho\{ \tilde s, O^{-1}\pd{O}{\tilde{\bf q}} \} -
\{ \tilde\rho, O^{-1}\pd{O}{\tilde{\bf q}} \}\tilde s ) = 0,
\end{equation}
so that the spin density remains gauge covariant. The cancellation
remains true for all orders in the expansion. Similarly, the charge
and current densities do not acquire
additional terms under the local gauge transformation introduced
above. The extra terms appearing as a result of the transformation
cancel when the trace is taken. When the change in the Wigner
function under a gauge transformation is taken into account the
overall expressions are gauge covariant.

\subsection{Gauge-covariant expressions for spin and torque dipoles}
Equation (\ref{ndip}) for the particle number density can be
formally written the following way
\begin{equation}
\label{nmod} n ({\bf R}, t) = {\rm tr} \int d^3 q\, \rho({\bf q},
{\bf R} - \bar{\cal R}).
\end{equation}
Re-expressing the integrand in this manner is tantamount to
making, in the density matrix, the replacement $\rho({\bf q}, {\bf
r}) \rightarrow \rho({\bf q}, {\bf r} - \bar{\cal R})$. To ensure
the gauge covariance of the number density the Fourier dual {\bf
r} is replaced by the true position ${\bf r} + \bar{\cal R}$, as
illustrated in Fig.~3. This is the same position as found in
section II. If the subspace contains one band expression
(\ref{nmod}) is an exact result, not simply a formal way of
writing the number density.

\begin{figure}[tbp]
\centering \epsfig{file=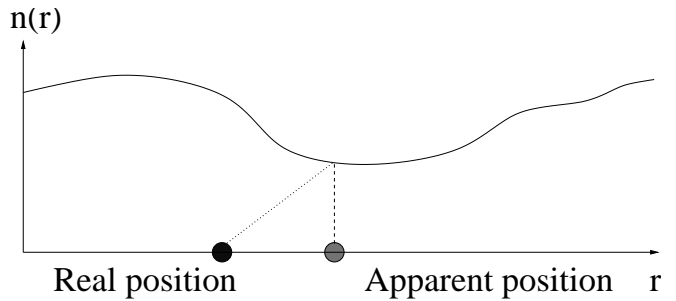, width=\columnwidth}
\caption{The position which appears in the Wigner function is not
the real position.} \label{fig:positionshift}
\end{figure}

Examining more closely the spin density as given in (\ref{s}) it
is evident that, whereas the spin density itself is gauge
covariant, its individual constituents are not. If one were to
consider the integrand of the dipole term in (\ref{s}) without the
Wigner function, this quantity would not by itself be gauge
covariant. We have already constructed a gauge-covariant spin
dipole $\bar{\bf p}^s = \bar{\bf M}^s - \frac{1}{2} \{ \bar{\cal
R}, \bar s \} $. In terms of the gauge-covariant spin dipole, the
spin density can be re-expressed as
\begin{equation}
\begin{split}
S ({\bf R}, t) = {\rm tr} \int \frac{d^3q}{(2\pi)^3} \, (\rho\,\bar s - \nabla_{\bf R} \cdot \rho\,\bar{\cal R} \, \bar s
 - \nabla_{\bf R} \cdot \rho\,\bar{\bf p}^s ) .
\end{split}
\end{equation}
It can be formally written the following way
\begin{equation}\label{sgc}
S({\bf R}, t) =  {\rm tr} \int \frac{d^3q}{(2\pi)^3}
\, [\rho({\bf q}, {\bf R} - \bar{\cal R})\, \bar s
- \nabla_{\bf
R} \cdot \rho({\bf q}, {\bf R} - \bar{\cal R})\,\bar{\bf p}^s].
\end{equation}
The gauge-covariant spin dipole $\bar{\bf p}^s$, as pointed out
above, is the result of a carrier's center of spin being different
from its center of charge.

In the same way the gauge-covariant torque dipole has been defined
by $ \bar{\bf p}^\tau = \bar{\bf M}^\tau - \frac{1}{2}\{ \bar{\cal
R}, \bar\tau \}$. Carrying out an identical manipulation to that
for the spin density, use of the gauge-covariant torque dipole
allows us to rewrite the torque density formally as
\begin{equation}
{\cal T}({\bf R}, t) = {\rm tr} \int \frac{d^3q}{(2\pi)^3} \,
[\rho({\bf q}, {\bf R} - \bar{\cal R})\, \bar\tau - \nabla_{\bf
R} \cdot \rho({\bf q}, {\bf R} - \bar{\cal R})\, \bar{\bf p}^\tau].
\end{equation}
A similar formal expression exists for the spin current.
Evidently, since these expressions are simply a rewriting of the
spin, torque and current densities, the equation of continuity
satisfied by them is unaltered.

We remark that the gauge-covariant quantities introduced
throughout this paper are well defined. In fact, even though the
density matrix was expressed in a basis of extended Bloch states,
by means of various manipulations we have been able to obtain
well-defined formulas for all the objects relevant in transport.

Finally, as described by Dudarev {\it et al.} \cite{Artem} in the
context of atomic physics, optical lattices can be constructed
which mimic the spin-orbit interaction. In such a lattice, a wave
packet can be constructed using cold atoms and the evolution of
its centers of mass and spin can be followed. In the same spirit,
Kato {\it et al.} \cite{Kato} have shown that it is possible to
follow the motion of a spin packet in solid state systems, which
demonstrates the feasibility of an analogous experiment in InGaAs.
This would provide a way to measure a spin dipole directly.

\section{Summary}
We have formulated a theory suitable for describing the dynamics
of particles in both single and multiple bands. The one-band
results known from the wave-packet formalism, including the terms
connected to the Berry phase, emerge from our theory. The
formalism can be applied to any clean system regardless of the
dimensionality of the Hilbert space under consideration. As a
result of the gauge degree of freedom of the basis functions, the
position vector which is used as a label of the carrier must be
modified by a gauge-dependent shift in order to obtain the true
particle position. We have shown the way to define macroscopic
density distributions for conserved and non-conserved operators.
In the case of spin we have highlighted the correspondence between
the multipole terms which appear in the density-matrix formalism
and those found earlier in the wave-packet formalism and we have
shown that experiments can be performed to identify the effect of
a spin dipole.

DC was supported by the NSF under grant number DMR-0404252. QN was
supported by the DOE under grant number DE-FG03-02ER45958.

\section{Appendix} We will present in this appendix some of the
proofs and evaluations which require lengthier calculations and
would interrupt the flow if incorporated into the main text. For
all derivations except the first, the $\ket{\bar u_m}$ and
$\ket{u_m}$ are interchangeable.
\subsection{Proof of Eq. (\ref{nonadb}) for degenerate bands}
The perturbed eigenfunctions are given by Eq.(\ref{nonadb}):
\begin{equation}
\nonumber \ket{\bar u_{m}} = \ket{u_{m}} - \sum_{n\ne
m}\frac{\tbkt{u_{n }}{i\hbar\td{}{t}}{u_{m}}}{\varepsilon_m -
\varepsilon_n}\ket{u_{n}}.
\end{equation}
For a set of degenerate bands the equilibrium part of the density
matrix is proportional to the identity matrix $I$ and will have
the form $f_0 \, I$, where $f_0$ is usually the Fermi-Dirac
distribution. Transport theory often distinguishes between
intrinsic effects, which are due to the equilibrium part of the
density matrix, and extrinsic effects, which are due to the
non-equilibrium correction to it. If we wish to evaluate the
expectation value of an operator $\hat A$ using the equilibrium
part of the density matrix for a set of degenerate bands, which
for simplicity here we will take as being two dimensional, the
following quantity must be evaluated:
\begin{equation}\nonumber
\bkt{\hat A} = f_0(\tbkt{\bar u_1}{\hat A}{\bar u_1} + \tbkt{\bar
u_2}{\hat A}{\bar u_2})
\end{equation}
The perturbed wave functions are given by:
\begin{equation}\nonumber
\arraycolsep 0.3ex
\begin{array}{rl}
& \displaystyle \ket{\bar u_1} = \ket{u_1} - \frac{\tbkt{u_{2
}}{i\hbar\td{}{t}}{u_{1}}}{\varepsilon_1 -
\varepsilon_2}\ket{u_{2}} - \sum_{n\ne 1,2}\frac{\tbkt{u_{n
}}{i\hbar\td{}{t}}{u_{1}}}{\varepsilon_m -
\varepsilon_n}\ket{u_{n}}\\ [2.2 ex] & \displaystyle  \ket{\bar
u_{2}} = \ket{u_{2}} - \frac{\tbkt{u_{1
}}{i\hbar\td{}{t}}{u_{2}}}{\varepsilon_2 -
\varepsilon_1}\ket{u_{1}} - \sum_{n\ne 1,2}\frac{\tbkt{u_{n
}}{i\hbar\td{}{t}}{u_{m}}}{\varepsilon_m -
\varepsilon_n}\ket{u_{n}}.
\end{array}
\end{equation}
The expectation values are:
\begin{equation}\nonumber
\arraycolsep 0.3ex
\begin{array}{rl}
& \displaystyle \tbkt{\bar u_1}{\hat A}{\bar u_1} = A_{11} -
\frac{\dbkt{u_{1 }}{i\hbar\td{u_{2}}{t}}}{\varepsilon_1 -
\varepsilon_2}A_{21} - \frac{\dbkt{u_{2
}}{i\hbar\td{u_{1}}{t}}}{\varepsilon_1 - \varepsilon_2}A_{12} +
\sum_1^{out}
\\ [2.2 ex] & \displaystyle
\tbkt{\bar u_2}{\hat A}{\bar u_2} = A_{22} - \frac{\dbkt{u_{1
}}{i\hbar\td{u_{2}}{t}}}{\varepsilon_2 - \varepsilon_1}A_{21} -
\frac{\dbkt{u_{2 }}{i\hbar\td{u_{1}}{t}}}{\varepsilon_2 -
\varepsilon_1}A_{12} + \sum_2^{out}.
\end{array}
\end{equation}
In the above $\sum_1^{out}$ and $\sum_2^{out}$ stand for the sums
involving the bands outside the degenerate manifold and $A_{ij} =
\tbkt{u_i}{\hat A}{u_j}$. The denominators of the terms involving $A_{12}$ and $A_{21}$ have opposite signs. Therefore, adding up the expectation values
we obtain:
\begin{equation}\nonumber
(\tbkt{\bar u_1}{\hat A}{\bar u_1} + \tbkt{\bar u_2}{\hat A}{\bar
u_2})= A_{11} + A_{22} + \sum_1^{out} + \sum_2^{out}.
\end{equation}
Therefore the terms with diverging denominators cancel out.

\subsection{Evaluation of position matrix elements}

In the general case, the expectation value of the position
operator is given by
\begin{equation}
\begin{split}
\nonumber
\bkt{\hat {\bf r}} = \int d{\cal V}\, \rho_{nn'} \int
d^3Q \, e^{-i{\bf Q}\cdot {\bf r}}\tbkt{\bar u_{n'-}}{ \hat {\bf r} e^{i{\bf Q}\cdot
\hat{\bf r} } }{\bar u_{n+}} \\
= \int d{\cal V}\, \rho_{nn'} \int d^3Q \, e^{-i{\bf Q}\cdot {\bf r}} \tbkt{\bar u_{n'-}}{(-i\pd{}{{\bf Q}} e^{i{\bf
Q}\cdot\hat{\bf r} }) }{\bar u_{n+}}.
\end{split}
\end{equation}
We can expand the integrand in the above equation and obtain
\begin{equation}\nonumber
\arraycolsep 0.3ex
\begin{array}{rl}
\tbkt{\bar u_{n'-}}{\hat {\bf
r} e^{i{\bf Q}\cdot\hat{\bf r} } }{\bar u_{n+}} = & \displaystyle - i\pd{}{{\bf Q}} \tbkt{\bar
u_{n'-}}{e^{i{\bf Q}\cdot\hat{\bf r}} }{\bar u_{n+}}  \\ [2.2ex] & \displaystyle +
\tbkt{i\pd{\bar u_{n'-}}{{\bf Q}}}{e^{i{\bf Q}\cdot\hat{\bf r}}
}{\bar u_{n+}}  \\ [2.2ex] & \displaystyle + \tbkt{\bar u_{n'-}}{e^{i{\bf Q}\cdot\hat{\bf r}}
}{i\pd{\bar u_{n+}}{{\bf Q}}}  \\ [2.2ex] & \displaystyle + {\bf r}\tbkt{\bar
u_{n'-}}{e^{i{\bf Q}\cdot\hat{\bf r}} }{\bar u_{n+}}.
\end{array}
\end{equation}
All four brackets represent integrals of products of lattice
periodic functions and exponentials. They will all eventually be
proportional to $\delta({\bf Q})$. Because of this fact the first
term above integrates to zero. The partial derivatives in the
second and third terms are evaluated by expanding the lattice
periodic Bloch wave functions about {\bf q}, treating {\bf Q} as a
small parameter
\begin{equation}
\begin{split}\label{uexpand}
\ket{\bar u_{n+}} = \ket{\bar u_{n\tilde{\bf q}}} + \frac{{\bf
Q}}{2}\cdot\ket{\pd{\bar u_{n\tilde{\bf q}}}{\tilde{\bf q}}}
\\
\ket{\bar u_{n-}} = \ket{\bar u_{n\tilde{\bf q}}} - \frac{{\bf
Q}}{2}\cdot\ket{\pd{\bar u_{n\tilde{\bf q}}}{\tilde{\bf q}}}
\end{split}
\end{equation}
In the limit in which ${\bf Q}\rightarrow 0$
\begin{equation}\nonumber
\ket{\pd{\bar u_{n\pm}}{{\bf Q}}} = \pm\frac{1}{2}\ket{\pd{\bar
u_{n\tilde{\bf q}}}{\tilde{\bf q}}}.
\end{equation}
Consequently
\begin{equation}\nonumber
\tbkt{\bar u_{n'-}}{\hat {\bf
r} e^{i{\bf Q}\cdot\hat{\bf r} } }{\bar u_{n+}} = {\bf r}\delta_{n'n} + \bar{\cal R}_{n'n},
\end{equation}
and the expectation value of the position vector yields, finally
\begin{equation}
\begin{split}\nonumber
\bkt{\hat{\bf r}} = \int d{\cal V}\,
\rho_{nn'}({\bf r}\delta_{n'n} + \bar{\cal R}_{n'n}).
\end{split}
\end{equation}

\subsection{Evaluation of spin and torque gauge fields}
The spin gauge field is found in an analogous fashion to the
expectation value of the position operator
\begin{equation}
\begin{split}
\nonumber \bkt{\hat {\bf r}\hat s} = \int d{\cal V}\, \rho_{nn'}
\int d^3Q \, e^{-i{\bf Q}\cdot {\bf r}}\tbkt{\bar u_{n'-}}{ \hat
{\bf r} \hat s e^{i{\bf Q}\cdot
\hat{\bf r} } }{\bar u_{n+}} \\
= \int d{\cal V}\, \rho_{nn'} \int d^3Q \, e^{-i{\bf Q}\cdot {\bf
r}} \tbkt{\bar u_{n'-}}{\hat s (-i\pd{}{{\bf Q}} e^{i{\bf
Q}\cdot\hat{\bf r} }) }{\bar u_{n+}} \\
= \int d{\cal V}\, \rho_{nn'} \bar{\bf M}^s_{n'n}.
\end{split}
\end{equation}
Expanding the above,
\begin{equation}\nonumber
\arraycolsep 0.3ex
\begin{array}{rl}
\tbkt{\bar u_{n'-}}{\hat {\bf r}\hat s e^{i{\bf Q}\cdot\hat{\bf r}
} }{\bar u_{n+}} = & \displaystyle - i\pd{}{{\bf Q}} \tbkt{\bar
u_{n'-}}{\hat s e^{i{\bf Q}\cdot\hat{\bf r}} }{\bar u_{n+}}  \\
[2.2ex] & \displaystyle + \tbkt{i\pd{\bar u_{n'-}}{{\bf Q}}}{\hat
s e^{i{\bf Q}\cdot\hat{\bf r}} }{\bar u_{n+}}  \\ [2.2ex] &
\displaystyle + \tbkt{\bar u_{n'-}}{\hat s e^{i{\bf
Q}\cdot\hat{\bf r}} }{i\pd{\bar u_{n+}}{{\bf Q}}} .
\end{array}
\end{equation}
Using the same arguments as in the previous section, the first
term integrates to zero, and after evaluating the brackets
involving the exponential and the lattice-periodic functions, we
obtain in the limit ${\bf Q} \rightarrow 0$
\begin{equation}
\bar{\bf M}^s_{n'n} = \frac{i}{2} (\tbkt{\bar u_{n'}}{\hat
{s}}{\pd{\bar u_n}{{\bf q}}} - \tbkt{\pd{\bar u_{n'}}{{\bf
q}}}{\hat {s}}{\bar u_{n}}).
\end{equation}

An almost identical derivation applies for the expectation value
$\bkt{\hat{\bf r}\hat{\tau}} \equiv \frac{1}{2}\bkt{\{\hat{\bf r},
\hat{\tau}\}}$. The only difference is that, since $\hat\tau$ may
be a function of wave vector, an additional term involving $-
i\pd{\hat{\tau}}{{\bf Q}}$ is generated. However, that term drops
out when the anti-commutator is taken, leaving us with the result
\begin{equation}
\bar{\bf M}^\tau_{n'n} = \frac{i}{2} (\tbkt{\bar u_{n'}}{\hat
{\tau}}{\pd{\bar u_n}{{\bf q}}} - \tbkt{\pd{\bar u_{n'}}{{\bf
q}}}{\hat {\tau}}{\bar u_{n}}).
\end{equation}

\subsection{Finite variance of carrier position}

Since the expectation value of the carrier position has been shown above
to contain no divergences, if the expectation value $\bkt{\hat {\bf r}^2}$ is
not divergent then the variance $\bkt{\hat {\bf r}^2} - \bkt{\hat
{\bf r}}^2$ is also finite. To see that the variance of the particle position does not contain
any divergences, we need to evaluate the expectation value
$\bkt{\hat {\bf r}^2}$. This requires us to evaluate the
matrix element
\begin{equation}
\begin{split}\nonumber
\tbkt{\bar u_{n'-}}{\hat
r^2 e^{i{\bf Q}\cdot \hat{\bf r}}}{\bar u_{n+}} = - \tbkt{\bar u_{n'-}}{( \frac{\partial^2}{\partial
Q^2}e^{i{\bf Q}\cdot \hat{\bf r}})}{\bar u_{n+}}.
\end{split}
\end{equation}
First note that an expression of the
form $-A \frac{\partial^2 B}{\partial Q^2} CD$ can be written as
\begin{equation}\nonumber
\arraycolsep 0.3ex
\begin{array}{rl}
-A \frac{\partial^2 B}{\partial Q^2} CD = &\displaystyle- \frac{\partial^2
}{\partial Q^2}(ABCD) \\ [2.2ex] &\displaystyle + 2\pd{}{{\bf Q}}\cdot[\pd{A}{{\bf Q}} BCD +
 AB \pd{}{{\bf Q}}(CD)] \\ [2.2ex]  &\displaystyle - \frac{\partial^2 A}{\partial Q^2}BCD - 2\pd{A}{{\bf Q}} \cdot B\pd{}{{\bf Q}}(CD) \\ [2.2ex] &\displaystyle - AB \frac{\partial^2
B}{\partial Q^2} (CD).
\end{array}
\end{equation}
In the case we are considering, all the products involve brackets
which are proportional to $\delta({\bf Q})$. As a result, all the
terms on the first line vanish under integration with respect to
{\bf Q} and only the terms on the second line need to be
evaluated. Writing them out explicitly,
\begin{equation}\nonumber
\arraycolsep 0.3ex
\begin{array}{rl}
&\displaystyle \tbkt{\bar u_{n'-}}{\hat
r^2 e^{i{\bf Q}\cdot \hat{\bf r}}}{\bar u_{n+}}
= r^2 \tbkt{\bar u_{n'-}}{e^{i{\bf
Q}\cdot\hat{\bf r}}}{\bar u_{n+}}) \\[2.2ex] &\displaystyle - \tbkt{\frac{\partial^2 \bar
u_{n'-}}{\partial Q^2}}{e^{i{\bf Q}\cdot\hat{\bf r}}}{\bar u_{n+}}
- \tbkt{\bar u_{n'-}}{e^{i{\bf Q}\cdot\hat{\bf
r}}}{\frac{\partial^2 \bar u_{n+}}{\partial Q^2}}
\\[2.2ex] &\displaystyle
+ 2i{\bf r}\cdot(\tbkt{\pd{\bar u_{n'-}}{{\bf Q}}}{e^{i{\bf
Q}\cdot\hat{\bf r}}}{\bar u_{n+}} + \tbkt{\bar u_{n'-}}{e^{i{\bf
Q}\cdot\hat{\bf r}}}{\pd{\bar u_{n+}}{{\bf Q}}})\\[2.2ex] &\displaystyle - 2\tbkt{\pd{\bar
u_{n'-}}{{\bf Q}}}{e^{i{\bf Q}\cdot\hat{\bf r}}}{\pd{\bar
u_{n+}}{{\bf Q}}}.
\end{array}
\end{equation}

In order to evaluate the differentials, the wave functions are
expanded as in (\ref{uexpand}) except that now the expansion must
be made to second order in {\bf Q}. The final result is
\begin{equation}\nonumber
\arraycolsep 0.3ex \begin{array}{rl} \bkt{\hat {\bf r}^2} = &
\displaystyle \int d{\cal V}\, \rho_{nn'}[r^2\delta_{n'n} + 2{\bf
r}\cdot\bar{\cal R}_{n'n} \\ [2.2ex] & \displaystyle +
\frac{1}{2}\bra{\pd{\bar u_{n'}}{\tilde{\bf q}}}\cdot\ket{\pd{\bar
u_{n}}{\tilde{\bf q}}} - \frac{1}{4}(\dbkt{\frac{\partial^2\bar
u_{n'}}{\partial \tilde q^2}}{\bar u_n} \\ [2.2ex] & \displaystyle
+ \dbkt{\bar u_{n'}}{\frac{\partial^2 \bar u_{n}}{\partial \tilde
q^2}}) ].
\end{array}
\end{equation}
This is clearly finite so the variance of the position operator is finite.

\subsection{Evaluation of velocity matrix elements}

The velocity matrix element $\tbkt{\bar u_{n'-}}{ e^{-i{\bf
q}_-\cdot\hat{\bf r} } \hat {\bf v} e^{i{\bf q}_+\cdot \hat{\bf r}
} }{\bar u_{n+}}$ is easily evaluated as
\begin{equation}\nonumber
\arraycolsep 0.3 ex
\begin{array}{rl}
&\displaystyle\tbkt{\bar u_{n'-}}{ e^{-i{\bf q}_-\cdot\hat{\bf r}
} \hat {\bf v} e^{i{\bf q}_+\cdot \hat{\bf r} } }{\bar u_{n+}}  \\
[2.2 ex] &\displaystyle = \int d^3r' \, \bar u_{n'-} ({\bf r}')
e^{-i{\bf q}_-\cdot{\bf r}'} \hat {\bf v} e^{i{\bf q}_+\cdot {\bf
r}'} \bar u_{n+}({\bf r}')  \\ [2.2 ex] &\displaystyle = \int
d^3r' \, e^{i{\bf Q}\cdot {\bf r}'} \bar u_{n'-} ({\bf r}')
(e^{-i{\bf q}_+\cdot\hat{\bf r}'} \hat {\bf v} e^{i{\bf
q}_+\cdot\hat{\bf r}'}) \bar u_{n+}({\bf r}')  \\[2.2 ex] &\displaystyle  =\delta({\bf
Q})\int d^3r' \, \bar u_{n'} ({\bf r}') \tilde {\bf v} \bar
u_{n}({\bf r}') = \delta({\bf Q})\tbkt{\bar u_{n'}}{\tilde {\bf
v}}{\bar u_{n}}.
\end{array}
\end{equation}
In the above we have abbreviated $\tilde {\bf v} = e^{-i{\bf q}\cdot\hat{\bf r}'} \hat {\bf v} e^{i{\bf q}\cdot\hat{\bf r}'} $.

We also need to evaluate the commutator $\frac{1}{2}\tbkt{\bar
u_{n'-}}{ e^{-i{\bf q}_-\cdot (\hat{\bf r} - {\bf r}) } \{ \hat
{\bf v}, (\hat {\bf r} - {\bf r})\} e^{i{\bf q}_+ \cdot (\hat{\bf
r} - {\bf r}) } }{\bar u_{n+}}$. One half of the commutator is
evaluated as
\begin{equation}\nonumber\arraycolsep 0.3 ex
\begin{array}{rl}
&\displaystyle\tbkt{\bar u_{n'-}}{ e^{-i{\bf q}_-\cdot (\hat{\bf
r} - {\bf r}) } \hat {\bf v}(\hat {\bf r} - {\bf r}) e^{i{\bf
q}_+\cdot (\hat{\bf
r} - {\bf r}) } }{\bar u_{n+}}  \\[2.2 ex]  &\displaystyle = \tbkt{\bar u_{n'-}}{ e^{-i{\bf
q}_-\cdot (\hat{\bf r} - {\bf r}) } \hat {\bf v}[-i\pd{}{{\bf
q}_+} e^{i{\bf q}_+\cdot (\hat{\bf r} - {\bf r}) } ]}{\bar u_{n+}}
\\[2.2 ex]
& \displaystyle = -i\pd{}{{\bf q}_+} \tbkt{\bar u_{n'-}}{
e^{-i{\bf q}_-\cdot (\hat{\bf r} - {\bf r}) } \hat {\bf v}
e^{i{\bf q}_+\cdot
(\hat{\bf r} - {\bf r}) } }{\bar u_{n+}} \\[2.2 ex] &\displaystyle + \tbkt{\bar u_{n'-}}{
e^{-i{\bf q}_-\cdot (\hat{\bf r} - {\bf r}) } \pd{\hat {\bf
v}}{{\bf q}_+} e^{i{\bf q}_+\cdot (\hat{\bf r} - {\bf r}) } }{\bar
u_{n+}} \\ [2.2 ex] &\displaystyle + \tbkt{\bar u_{n'-}}{ e^{-i{\bf q}_-\cdot (\hat{\bf r} -
{\bf r}) } \hat {\bf v} e^{i{\bf q}_+\cdot (\hat{\bf r} - {\bf r})
} }{i\pd{\bar u_{n+}}{{\bf q}_+}},
\end{array}
\end{equation}
while the other is
\begin{equation}\nonumber\arraycolsep 0.3 ex
\begin{array}{rl}
&\displaystyle\tbkt{\bar u_{n'-}}{ e^{-i{\bf q}_-\cdot (\hat{\bf r} - {\bf r}) }
(\hat {\bf r} - {\bf r}) \hat {\bf v} e^{i{\bf q}_+\cdot (\hat{\bf
r} - {\bf r}) } }{\bar u_{n+}} \\[2.2 ex] &\displaystyle  = \tbkt{\bar u_{n'-}}{ [i\pd{}{{\bf
q}_-} e^{-i{\bf q}_-\cdot (\hat{\bf r} - {\bf r}) } ] \hat {\bf v}
e^{i{\bf q}_+\cdot (\hat{\bf r} - {\bf r}) } }{\bar u_{n+}}
\\[2.2 ex] &\displaystyle
= i\pd{}{{\bf q}_-} \tbkt{\bar u_{n'-}}{ e^{-i{\bf q}_-\cdot
(\hat{\bf r} - {\bf r}) } \hat {\bf v} e^{i{\bf q}_+\cdot
(\hat{\bf r} - {\bf r}) } }{\bar u_{n+}} \\[2.2 ex] &\displaystyle  - \tbkt{i\pd{\bar
u_{n-}}{{\bf q}_-}}{ e^{-i{\bf q}_-\cdot (\hat{\bf r} - {\bf r}) }
\hat {\bf v} e^{i{\bf q}_+\cdot (\hat{\bf r} - {\bf r}) } }{\bar
u_{n'+}} \\[2.2 ex] &\displaystyle  - \tbkt{\bar u_{n'-}}{ e^{-i{\bf q}_-\cdot (\hat{\bf r} -
{\bf r}) } \pd{\hat {\bf v}}{{\bf q}_-} e^{i{\bf q}_+\cdot
(\hat{\bf r} - {\bf r}) } }{\bar u_{n+}}.
\end{array}
\end{equation}
All the brackets in the above two equations are proportional to $\delta({\bf
Q})$, causing most of the terms to cancel. We are left with:
\begin{equation}
\begin{split}\nonumber
\frac{1}{2}\tbkt{\bar u_{n'-}}{ e^{-i{\bf q}_-\cdot (\hat{\bf r} -
{\bf r}) } \{ \tilde {\bf v}, (\hat {\bf r} - {\bf r})\} e^{i{\bf
q}_+ \cdot (\hat{\bf r} - {\bf r}) } }{\bar u_{n+}} \\
= \frac{i}{2}\delta({\bf Q})(\tbkt{\bar u_{n}}{ \tilde {\bf v}
}{\pd{\bar u_{n'}}{{\bf q}}} - \tbkt{\pd{\bar u_{n}}{{\bf q}}}{
\tilde {\bf v} }{\bar u_{n'}} ).
\end{split}
\end{equation}

\subsection{Effect of gauge transformation on $\rho({\bf q}, {\bf r})$}

The operator $\hat\rho$ must be invariant under gauge
transformations. From the expansion of the density operator in Bloch eigenstates,
\begin{equation}\nonumber
\hat \rho = \sum_{n,n'}\sum_{{\bf k}, {\bf k}'} \rho_{nn'} ({\bf
k}, {\bf k}') \ket{\bar\psi_{n\tilde{\bf
k}}}\bra{\bar\psi_{n'\tilde{\bf k}'}},
\end{equation}
it is evident that, if the wave functions change according to $
\ket{\bar\psi_{m\tilde{\bf k}}} \rightarrow O_{mn}(\tilde{\bf
k})\ket{\bar\psi_{n\tilde{\bf k}}} $, then $\rho_{nn'}({\bf k},
{\bf k}')$ must transform to $O^{-1}\rho O$. Using the definition
of $\rho_{nn'}({\bf k}, {\bf k}')$ in terms of the Wigner
function,
\begin{equation}\nonumber
\rho({\bf q}, {\bf r}) = \int \frac{d^3Q}{(2\pi)^3}\, e^{i{\bf
Q}\cdot{\bf r}}\rho ({\bf q}_+, {\bf q}_-),
\end{equation}
and remembering that ${\bf k} \equiv {\bf q}_+$ and ${\bf k}'
\equiv {\bf q}_-$, we obtain
\begin{equation}\nonumber
\rho({\bf q}, {\bf r}) \rightarrow \int\! \frac{d^3Q}{(2\pi)^3}
e^{i{\bf Q}\cdot{\bf r}} O^{-1}(\tilde{\bf q}_+) \rho ({\bf q}_+, {\bf
q}_-) O (\tilde{\bf q}_-).
\end{equation}
The matrices $O^{-1}(\tilde{\bf q}_+)$ and $O(\tilde{\bf q}_-)$ may be expanded
about their arguments as:
\begin{equation}\nonumber
\arraycolsep 0.3ex
\begin{array}{rl}
&\displaystyle O^{-1}(\tilde{\bf q}_+) = O^{-1}(\tilde{\bf q}) +
\frac{{\bf Q}}{2} \cdot \pd{O^{-1}(\tilde{\bf q})}{\tilde{\bf q}} + O(Q^2)
\\ [2.2ex]&\displaystyle
O(\tilde{\bf q}_-) = O(\tilde{\bf q}) - \frac{{\bf Q}}{2} \cdot \pd{O (\tilde{\bf
q})}{\tilde{\bf q}} + O(Q^2).
\end{array}
\end{equation}
Then, to first order in ${\bf Q}$, abbreviating $O (\tilde{\bf q})$ to
$O$ and $\rho ({\bf q}_+, {\bf q}_-)$ to $\rho_q$,
\begin{equation}
\nonumber\arraycolsep 0.3ex\begin{array}{rl}&\displaystyle O^{-1}(\tilde{\bf q}_+) \rho_q ({\bf q}_+, {\bf
q}_-) O (\tilde{\bf q}_-)
 \\[2.2ex]&\displaystyle
= O^{-1} \rho_q O  + \frac{{\bf Q}}{2} \cdot
(\pd{O^{-1}}{\tilde{\bf q}}\rho_q O - O^{-1} \rho_q
\pd{O}{\tilde{\bf q}})
\\[2.2ex]&\displaystyle
= \tilde\rho_q + \frac{{\bf Q}}{2} \cdot (\pd{O^{-1}}{\tilde{\bf q}}O
\tilde\rho_q - \tilde \rho_q O^{-1}\pd{O}{\tilde{\bf q}}),
\end{array}
\end{equation}
where $\tilde\rho_q = O^{-1} \rho_q O$. The last line is obtained
by inserting $OO^{-1}$ or $O^{-1}O$ as appropriate. Since
$\pd{}{\tilde{\bf q}} (OO^{-1}) = 0$, we have
$\pd{O^{-1}}{\tilde{\bf q}}O = - O^{-1}\pd{O}{\tilde{\bf q}}$ and
\begin{equation}
\begin{split}\nonumber
\rho \rightarrow \tilde\rho - \frac{{\bf Q}}{2} \cdot \{\tilde
\rho , O^{-1}\pd{O}{\tilde{\bf q}}\}.
\end{split}
\end{equation}
Finally, writing $- \frac{{\bf Q}}{2} = \frac{i}{2}\nabla_{\bf r}
e^{i{\bf Q}\cdot{\bf r}}$, we recover formula (\ref{rhogt}) for
the transformation of $\rho({\bf q}, {\bf r})$.

\end{document}